\documentclass[floatfix,aps,prd,showpacs,amsmath,
nofootinbib,preprintnumbers,twocolumn]{revtex4}

\usepackage{graphicx}

\newcommand{\beq}{\begin{equation}}
\newcommand{\eeq}{\end{equation}}
\newcommand{\bea}{\begin{eqnarray}}
\newcommand{\eea}{\end{eqnarray}}

\def\laq{~\raise 0.4ex\hbox{$<$}\kern -0.8em\lower 0.62
ex\hbox{$\sim$}~}
\def\gaq{~\raise 0.4ex\hbox{$>$}\kern -0.7em\lower 0.62
ex\hbox{$\sim$}~}

\def \r {\rho}

\def \Om {\Omega}

\begin{document}

\preprint{BA-TH/06-549}
\preprint{astro-ph/0610574}

\title{SNLS data are consistent with acceleration at $z\approx 3$}

\author{Luca Amendola$^{1}$, M. Gasperini$^{2,3}$ and Federico Piazza$^{4,5}$}

\affiliation{$^{1}$INAF/Osservatorio Astronomico di Roma, \\
 Via Frascati 33, 00040 Monteporzio Catone (Roma), Italy\\
 $^{2}$Dipartimento di Fisica, Universit\`{a} di Bari, Via G. Amendola
173, 70126 Bari, Italy\\
 $^{3}$INFN, Sezione di Bari, Bari, Italy\\
 $^{4}$ Perimeter Institute for Theoretical Physics, 
 31 Caroline St. N, Waterloo Ontario N2L 2Y5, Canada\\
$^{5}$Institute of Cosmology and Gravitation, Merchantile House\\
University of Portsmouth,  Portsmouth, PO1 2EG, UK}

\begin{abstract}
We point out that the Type Ia supernovae in the SNLS dataset are consistent with an early beginning of the cosmic acceleration if dark energy interacts strongly with dark matter. We find that the acceleration could have started at redshift $z\approx 3$ and higher. 
\end{abstract}


\maketitle


One of the most puzzling aspects of cosmic acceleration, detected
through the combined analysis of supernovae, CMB and large-scale structure \cite{1}, is that the acceleration itself seems to have started at a relatively recent epoch, thus enforcing the so-called coincidence problem \cite{2}. Indeed, in a spatially flat universe filled with pressureless conserved matter $\r_m$ and with a fluid of equation of state $w(z)$, the standard Friedman equation
reads 
\begin{equation}
H^{2}=H_{0}^{2}\left[\Omega_{m}(1+z)^{3}+(1-\Omega_{m})(1+z)^{3(1+\hat{w})}\right],
\label{eq:hub1}
\end{equation}
where $\Omega_{m}=\r_m/\r_c$ is the present matter density parameter, $z=(a_0/a)-1$ is the redshift parameter, and $\hat{w}(z)\ln(1+z) =\int_{0}^{z}w(z')dz'/(1+z')$. Using the Einstein equation  $3 \ddot a /a= - 4 \pi G( \r+3 p)$, one then finds that the redshift at the beginning of the acceleration epoch is given by the (implicit) formula
\begin{equation}
z_{\textrm{acc}}=\left[(3w(z_\textrm{acc})+1)(\Omega_{m}-1)/\Omega_{m}\right]^{-1/3\hat{w}(z_\textrm{acc})}-1.
\label{eq:zacc}
\end{equation}
The recent analysis  of the SNLS collaboration \cite{2b}, illustrated in Fig. 1 (upper panel), has found that for a constant $w$ the values of $\Om_m$ and $w$ allowed by present observations on supernovae Ia and large-scale structure are $\Omega_{m}\in(0.2-0.4)$ and $w\in(-0.8,-1.2)$. One then obtains from Eq. (\ref{eq:zacc}) $z_{\textrm{acc}}\in (0.4-1)$ (see Fig. 1, lower panel). If $w$ is not a constant the result $z_{\textrm{acc}}\laq1$ is still generally valid,  unless $w(z)$ is wildly varying in recent epochs. 

\begin{figure}
\label{f1}
\begin{center}
\includegraphics[scale=0.55]{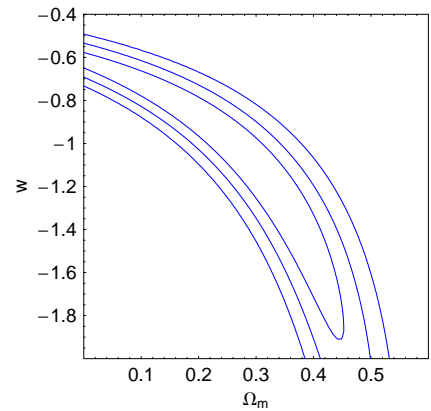}\\ 
\bigskip
\includegraphics[scale=0.8]{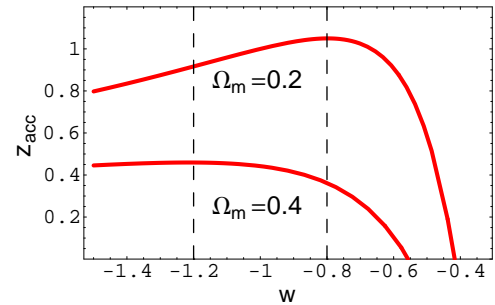}
\end{center}
\caption{\sl The upper panel shows the $w-\Omega_{m}$ confidence level for the SNLS dataset. The lower panel shows the beginning of the accelerated epoch for constant $w$, and for the two limiting values of $\Om_m$, according to Eq. (\ref{eq:zacc}).}
 \vspace{-.6cm}
\end{figure}

The aim of this note is to point out that a different interpretation
of Eq. (\ref{eq:hub1}) leads to quite a different conclusion on $z_{\textrm{acc}}$. Eq. (\ref{eq:hub1}) is based on the separate covariant conservation of the (dust) matter fluid and of the dark-energy fluid in Einsteinian gravity. Let us assume instead that the dark-matter component of $\Om_m$ (but not the baryons) interacts strongly with dark energy, so that the two ``dark" components of the cosmic fluid 
follow a scaling regime in which $\rho_{DM}\sim\rho_{DE}\sim a^{-3(1+w_{\textrm{eff}})}$ (where $w_{\textrm{eff}}$ is a constant, for simplicity), while the baryonic component follows the standard scaling behaviour of dust matter, $\r_B \sim a^{-3}$. The sum of dark matter and dark energy thus behaves as a single fluid, with effective barotropic parameter $w_{\textrm{eff}}$ and density parameter $\Omega_{DM}+\Omega_{DE}=1-\Omega_{b}$, where $\Omega_{b}= \r_B/\r_c$ is the baryonic density parameter. In this case, Eqs. (\ref{eq:hub1}-\ref{eq:zacc}) can be rewritten as 
\begin{equation}
H^{2}  =  H_{0}^{2}\left[\Omega_{b}(1+z)^{3}+(1-\Omega_{b})(1+z)^{3(1+w_{\textrm{eff}})}\right],
\label{eq:hub2}
\end{equation} 
\vspace{-.5cm}
\begin{equation}
z_{\textrm{acc}}(\Omega_{b},w_{\textrm{eff}})  =  \left[(3w_{\textrm{eff}}+1)(\Omega_{b}-1)/\Omega_{b}\right]^{-1/3w_{\textrm{eff}}}-1.
\label{eq:zacc2}
\end{equation}

The supernovae results of Fig. 1 can now be applied to Eq. (\ref{eq:hub2}), and reinterpreted in terms of $\Omega_{b}$ 
and $w_{\textrm{eff}}$  (the interpretation of the large-scale data requires  of course a model of the clustering growth that is beyond the scope of this short note). Reading the values $\Omega_{b}\approx0.04-0.05$ \cite{2b} we obtain $w_{\textrm{eff}}\approx-0.65\pm0.15$. Eq. (\ref{eq:zacc2}) then implies $z_{\textrm{acc}}\approx3-4$ (see Fig. 2). Therefore, a beginning of the acceleration at epochs much earlier than suggested by the commonly used interpretation of the data is consistent with SNLS results. We note that this new interpretation may also alleviate the need of ``phantom" dark energy \cite{2c} with ``supernegative" equation of state. 

\begin{figure}[t]
\includegraphics[bb=50bp 550bp 598bp 760bp,clip,scale=0.85]{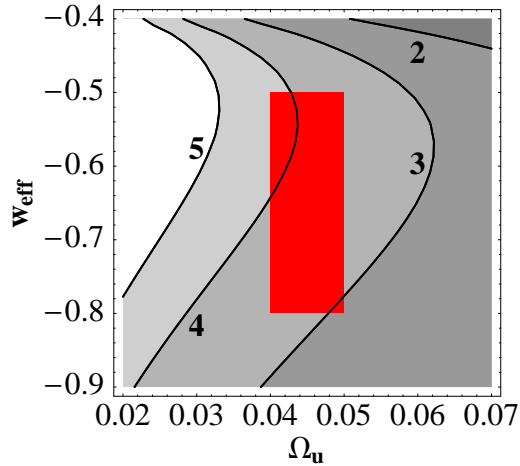}
\caption{\sl The allowed region of the $\{\Omega_{b},w_{\textrm{eff}}\}$ plane (the red rectangle) is superimposed to the  curves of constant $z_{acc}$, plotted for $z_{\textrm{acc}}=2,3,4,5$. }
\end{figure}

The scaling regime $\rho_{DM}\sim\rho_{DE}\sim a^{-3(1+w_{\textrm{eff}})}$ can be achieved if dark matter and dark energy interact strongly. There are several models studied in literature that realize this kind of coupling (see \cite{3,4} for phenomenological models, and \cite{4a} for dilaton models in ``late-time" string-cosmology scenarios). The simplest modeling of this effect is in terms of a scalar field with exponential potential and (nonminimal) gravitational couplings to dark matter of order unity (in gravitational units); for generalization to Lagrangians with higher-order kinetic terms see Ref. \cite{PT}. Using the fluido-dynamical formalism the equations describing this coupling can be written as \vspace{-.3cm}
\begin{eqnarray}
\dot{\rho}_{DM}+3H\rho_{DM} & = & \delta\\
\dot{\rho}_{DE}+3H\rho_{DE}(1+w_{DE}) & = & -\delta
\label{4}
\end{eqnarray}
 
\vspace{-.2cm}
where $\delta=-3Hw_{\textrm{eff}}\rho_{DM}= 3H \r_{DE}(w_{\textrm{eff}}-w_{DE})$ (in the latter equality we neglected the baryons). Such a strong coupling might be in conflict with observations like CMB and large-scale structure, but in principle one can envisage a model with a variable coupling, and pass the constraints at early times (as in \cite{4b}). We should also note that, if dark energy behaves as a fluidodynamical medium with sound speed $c_{s}=1$ (as it occurs for a scalar field with standard kinetic term), its linear perturbations are much weaker than those of the dark-matter fluid. Indeed, although the two dark components behave as a single fluid at the (unperturbed)  background level, their clustering properties are clearly very different. 

It should be mentioned, finally, that the arguments presented here can  be generalized to the case in which only part of  dark matter is coupled to dark energy, as proposed in \cite{paper1}. In that case one must replace $\Om_b$ with $\Om_u$ in Eqs.  (\ref{eq:hub2}-\ref{eq:zacc2}), 
where  $\Omega_{u}$ is defined as the sum of all the uncoupled
components. Using the "gold" data sample of \cite{9} we found in our previous paper \cite{paper1} that  $\Omega_{u}$ had to be bigger than $0.2$, roughly, to be consistent at 2$\sigma$ with an early beginning of the accelerated epoch (more precisely, with $z_{acc} \approx 3$). As has been shown here, instead, the new SNLS data \cite{2b} are consistent with an earlier acceleration even in the case in which only baryons are uncoupled to dark energy, i.e. $\Omega_{u}=\Omega_{b}\approx0.05$. Also, even the limiting case $\Omega_{u}\simeq 0$ is contemplated, in principle; with a chameleon-type \cite{cham} mechanism, in fact, even baryons could be strongly coupled to a scalar dark-energy field, and yet satisfy the current bounds imposed by gravitational tests of the equivalence principle. In this extreme case, the beginning of acceleration could in principle be pushed back to the end of the radiation era. 

{\it Acknowledgment.} We thank Elisabetta Majerotto for the drawing of Fig. 1, upper panel.

\end{document}